\documentstyle[12pt]{article}
\newcommand{\lam}{\lambda}
\newcommand{\be}{\begin{equation}}
\newcommand{\ee}{\end{equation}}
\newcommand{\ba}{\begin{eqnarray}}
\newcommand{\ea}{\end{eqnarray}}

\newcommand{\rar}{\rightarrow}
\begin{document}

\begin{center}
{\LARGE {\bf \rule{0em}{4em} Three-body dispersion-relation N/D
equations for the coupled decay channels \\
$\bar p p$ ($J^{PC}=0^{-+}$) $ \rightarrow \pi^0 \pi^0
\pi^0$, $\eta \pi^0 \pi^0$, $\eta \eta \pi^0$,  $\bar K K \pi^0$
}}\\ \vspace{0.8cm} A. V. Anisovich\\ Petersburg Nuclear Physics
Institute\\
Gatchina, St. Petersburg, 188350 Russia\\
e-mail: aanisovi@thd.pnpi.spb.ru\\\vspace{0.2cm}
{\bf Abstract} \\
\vspace{1.2em}
\parbox{30em}{During several years the data on different
channels $\bar p p $ ($J^{PC}=0^{-+}$) $\rightarrow 3\; mesons$
presented by Crystal Barrel Collaboration were  successfully
analyzed by extracting the leading amplitude singularities ---
 pole singularities ---
 with the aim to obtain information about two-meson
resonances.  But these analyses do not take into account
three-body final state interactions (FSI) in an explicitly correct
way.  This paper is devoted to the consideration of this problem. Here I
demonstrate how the coupled three-body equations may be written
for the $\pi^0 \pi^0 \pi^0,\;\; \eta \pi^0 \pi^0,\;\; \eta \eta
\pi^0, \;\; \bar K K \pi^0$ channels in the $\bar p p$ annihilation
at rest using three-body dispersion relation N/D-method.  } \\
\end{center}

\section{Introduction}

During several  years Crystal Barrel Collaboration presents the high
statistics data on three meson production from the $\bar p p$
annihilation at rest. These data were successfully analyzed
(see, for example, \cite{CB95,CB94,ABSZ94,AS96})  with the aim to
search for new meson resonances in the region
1000-1600 MeV using the K-matrix formalism or a simplified
dispersion
N/D-method. There is a strong expectation based on QCD
\cite{1011} as well as on lattice
calculations \cite{1213} that the lowest scalar glueball is located in
this region. Thus, the identification of  scalar resonances in
the mass region 800-2000 MeV and their classification in
$\bar q q$-nonets should be done to trap the lightest
scalar glueball. The quark/qluon structure of these resonances
may be determined from the analysis of coupling constants
 of these states to
 pseudoscalar mesons \cite{Gluonstructure}.  The question is: is the
 K-matrix approximation -- or simplified N/D-method --  sufficient for this
 purpose?

In this paper the three-body dispersion N/D-method is presented,
which is based on the two-body unitarity condition and allows one to take
into account FSI of three mesons.  The basic
principles of this technique were developed in 1960s and
they were applied to  the
calculation of FSI in the K-meson decay into three pions (see
\cite{Anis-Ansel66, Aitchison65} and references therein). Recently this
technique was used for the calculation of the $\eta \rightarrow
3 \pi$ decay \cite{Eta96} and $\phi$-meson production in the $p
\bar p$ annihilation at rest \cite{AK96}.

This paper is organized as follows. In Section 2 the
two-particle discontinuity of the decay amplitude is written
out, and the integral equation is derived. It is shown how to
 take into account not only the S-wave binary interactions but
 also interactions with higher angular momentum (P- and
 D-wave).  In Section 3 this technique is generalized for the
 resonance and non-resonance two-particle interactions.
 The complete three-body dispersion equation is written, that
 may be used for the analysis of  various three-particle
 reactions. In Section 4 it is shown how the coupled three-body
equations could be written for $\pi^0 \pi^0 \pi^0,\;\; \eta
\pi^0 \pi^0,\;\; \eta \eta \pi^0, \;\; \bar K K \pi^0$
channels in the $\bar p p$ annihilation from the $0^{-+}$ state.

\section{Two-particle discontinuity of the decay
amplitude}

\vspace{1cm}

Let us start with the consideration of
 the decay of a scalar particle with the mass
 $M$ and momentum $P$
into three scalar particles with masses $m_1$, $m_2$ and $m_3$
and momenta $k_1$, $k_2$ and $k_3$.
I would like
 to show that the energy dependence of this transition amplitude
may be derived from the unitarity and analyticity. The imaginary
part of the amplitude $A_n$ for the transition $1 \rightarrow n$ is given by
\be
Im M_n= \frac{1}{2} \sum_{n'} (2\pi)^4 \delta(P- \sum_{i=1}^{n'}
k_i) T^*_{n'n} M_{n'},
                                                       \label{u1}
\ee
where $T_{n'n}$ is the scattering matrix.
Equation (\ref{u1}) may be simplified if  the
approximation is used where $i$) binary interactions of particles are
taken into account, three-particle forces being neglected;
$ii$) there are no transitions through intermediate states
 with more than three particles. It is
 also supposed that there are  no transitions
through intermediate states involving other  particles; still, this is not
 a crucial restriction. Below it is shown how the production of
new particles in the intermediate state may be taken into
account.  In this approximation one
 may write eq.(\ref{u1}) as follows:
 \be Im
M_3= \frac{1}{2} (2\pi)^4 \delta(P-k_1-k_2-k_3) T^*_{33} M_{3}.
                                                 \label{u2}
\ee

Three-particle forces being neglected,
the amplitude $M_3$ for the three-particle production is
a sum of four terms: $i$) the direct production amplitude
 $\lam(s_{12},s_{13},s_{23})$, which is assumed, for the sake of
simplicity, to
be  free of singularities; $ii$) the amplitude $a_{12}$, where
the last interaction is that of particles 1 and 2;  likewise,
there are similar terms $iii$) $a_{13}$ and $iiii$) $a_{23}$.
I take into
account FSI with different orbital momenta $L$, hence
 the amplitude $a_{ij}$ is written as
  \be
  a_{ij}= \sum_L F_L
 A^L_{ij}(s_{ij}). \label{u3}
  \ee
  Here $s_{ij}$ is the two-particle invariant
mass squared, $s_{ij}=(k_i + k_j)^2$. The function $F_L$ defines
the angular distribution of decaying particles. In the case of
the S-wave binary interaction $F_0=1$, but for the higher-wave
interaction the amplitude depends on the angle between
interacting and spectator particles as well.

\subsection{ S-wave interaction}

Let us begin with the simplest case of the S-wave pair
interactions. The decaying amplitude is given by \be
M(s_{12},s_{13},s_{23})= \lam(s_{12},s_{13},s_{23}) + A^0_{12}(s_{12}) +
A^0_{13}(s_{13}) +A^0_{23}(s_{23}).
                                                        \label{u4}
\ee
The connected part of the matrix $T_{33}$ is not known, so I
cannot directly calculate the contribution from three-particle
intermediate state into the imaginary part of $M$ and write down
 corresponding dispersion integral. So, I will explore the
two-particle unitarity condition to derive the
integral equation  for the amplitude $A^0_{ij}$.
This method was described in details in ref.
\cite{Anis-Ansel66}. The idea of such an approach is that one should
consider the case of small external mass $M< m_1+m_2+m_3$, when
only the scattering reactions are physically possible. A
simple relation can be written in this case, and then I perform
analytical continuation of  the final equation over the mass $M$
back to the decay region. Let us stress that the obtained
expression  differs from the contribution from the disconnected
part of $T$ into imaginary part $M$: this result satisfies the
three-body unitarity condition, in which three particles
 interact by pairs only \cite{Aitchison66}.

So, I write down an ordinary unitarity condition for the
scattering in the channel of particles 1 and 2. This
means that $(M+m_3) < s_{12}$ is assumed.  The discontinuity  of
the amplitude is equal to \begin{eqnarray}
disc_{12}M(s_{12},s_{13},s_{23}) =
\nonumber
\end{eqnarray}
\begin{eqnarray}
= \frac{1}{2} \int d\Phi_{12}(k_1,k_2) \biggl( \lam
+ A^0_{12}(s_{12}) + A^0_{13}(s_{13}) +A^0_{23}(s_{23}) \biggr)
A_{2 \rightarrow 2}^{0*}(s_{12})   ,
                                                        \label{u5}
\end{eqnarray}
where
$d\Phi_{12}(k_1,k_2)$ is the phase volume of  particles 1
and 2
\be
d\Phi_{12}(k_1,k_2)= (2\pi)^4 \delta^4(P-k_1-k_2)\;\frac{d^4 k_1 d^4 k_2}
{(2\pi)^6} \delta(m_1^2-k_1^2) \delta(m_2^2-k_2^2).
\ee
 $A^0_{2\rightarrow 2}$ is the S-wave two-particle scattering
  amplitude,
 which  can be written in the dispersion N/D method  as
a series
\ba
A^0_{2\rightarrow 2}(s)= G_0^L(s) G_0^R(s) +
G_0^L(s) B_0(s) G_0^R(s)+ G_0^L(s) B_0^2(s) G_0^R(s)+\; ...\; =
\nonumber
\ea
 \be
 = \frac{G_0^L(s) G_0^R(s)}{1-B_0(s)}, \label{u6}
\ee
where $G_0^L$ and $G_0^R$ are left and right
vertex functions and $B(s)$ is the dispersion representation of
the loop diagram:
\be
B_0(s)= \int_{(m_i+m_j)^2}^{\infty} \frac{ds'}{\pi}
\frac{G_0^L(s') \rho_{ij}(s') G_0^R(s')}{s'-s},
\ee
where $\rho_{ij}$ is the two-particle phase space
\be
\rho_{ij}(s)= \frac{1}{16\pi s} \sqrt{[s-(m_1+m_2)][s-(m_1-m_2)]}\;.
\ee
Vertex functions contain left-hand singularities related to the
t-channel exchange diagrams, while B-function has singularities
due to the elastic scattering.
 It is not specified
 from the  consideration of the scattering
amplitude $A^0_{2\rightarrow 2}$ of eq. (\ref{u5})
whether both vertices $G_0^L$ and $G_0^R$ have these singularities
or only one of them. In the case of three-body decay the situation is
 quite opposite. On the first sheet the decay amplitude
 has  only singularities at
$s_{ij}= (m_i+m_j)^2$, which are associated with the elastic
scattering in the subchannel of particles $i$ and $j$. This
means that the vertex $G_0^R$ is analytical function. The
simplest choice is
 \be
  G_0^R=1
  \ee
Now return to eq. (\ref{u5}).
Since the functions $\lambda$ do not have the two-particle threshold
singularity and thus do not have
a discontinuity, the lhs of eq. (\ref{u5}) is
\be
disc_{12}M(s_{12},s_{13},s_{23}) = disc_{12}A^0_{12}(s_{12}).
                                                        \label{u7}
\ee
As was stressed in ref. \cite{Eta93},
 only one rescattering of particles 1 and 2 can be
considered in the final state
and a full set of binary rescatterings can be taken into account
 multiplying
by $(1-B_0(s_{12}))^{-1}$. Thus, the two-particle discontinuity
in this special case of one rescattering is defined as
\be
disc_{12} \; A^{0}_{12}(s_{12}) = \frac{1}{2} \int d\Phi_{12}(k_1,k_2)
\biggl( \lam(s_{12},s_{13},s_{23}) +
 A^0_{13}(s_{13}) +  A^0_{23}(s_{23})
 \biggr) G_0^L(s_{12}).
                                                        \label{u8}
\ee
It is convenient to perform the phase space integration in eq. (\ref{u8})
in the center-of-mass system of particles 1 and 2. In this frame
\ba
   s_{13}= m_1^2 +m_3^2- 2k_{10} k_{30} + 2z_{13}\mid \vec k_1\mid
   \mid \vec k_3\mid\;,
\nonumber\\
  k_{10} = \frac{s_{12}+m_1^2-m_2^2}{2\sqrt{s_{12}}}\;, \qquad
  \mid \vec k_1\mid=\sqrt{k_{10}^2-m_1^2},
   \label{u9}
\\
  k_{30} = \frac{s_{12}+m_3^2-s}{2\sqrt{s_{12}}}\;, \qquad
  \mid \vec k_3\mid=\sqrt{k_{30}^2-m_3^2},
\nonumber
\ea
where $z_{13}=\cos \theta_{13}$,
and $\theta_{13}$ is the angle between         particles 1 and 3
in the cms of particles 1 and 2, and $s=M^2$.
The expression for $s_{23}$ is obtained from eq. (\ref{u9}) by the
replacement $1 \leftrightarrow 2$. From eq. (\ref{u8}) one has:
\be
disc_{12}\; A^{0}_{12}(s_{12}) =
\biggl( \lam_S(s_{12}) +
\langle A^0_{13}(s_{13}) \rangle_0 +  \langle A^0_{23}(s_{23}) \rangle_0
 \biggr) G_0^L(s_{12}) \rho_{12}(s_{12}),
                                                        \label{u10}
\ee
where  the following notation is used:
\be
\langle A^0_{i3}(s_{i3})\rangle_0 =\;
 \int_{C_i(s_{12})} \frac{dz_{i3}}{2}\;A^0_{i3}(s_{i3})\;.
                                                          \nonumber
\ee
 $\lam_S$ is the S-wave projection of $\lam$:
\be
\lam_S(s_{12});=\;
\int_{-1}^{1} \frac{dz_{13}}{2}\;\lam(s_{12},s_{13},s_{23}).
\ee
Analytical continuation over external mass $M$ from the scattering
 to the decay region allows one to define correctly the rules of
integration over $z$.
This integration should be carried out along the contour
$C_i(s_{12})$, whose position at different $s_{12}$ is
 described in
detail in ref.\cite{Anis-Ansel66}. Here I would like to note that only
at small $s_{12}$,
\be
(m_1+m_2 )^2\;\leq s_{12} \leq
\frac{m_i s}{m_i+m_3} +\frac{m_3}{m_i+m_3}\;(m_1+
m_2-m_i)^2-m_i m_3 ,   \nonumber
\ee
it coincides with the phase space integration contour
\be
-1 \leq z_{i3} \leq 1,  \nonumber
\ee
and it contains an additional piece at larger $s_{12}$.

Equation (\ref{u10}) allows us to write down the dispersion integral for the
amplitude  with one pair rescattering in the final state:
 \be A^{0}_{12}(s_{12})= \lam_S(s_{12})
B(s_{12}) +  \int_{(m_1+m_2)^2}^{\infty}
 \frac{ds'_{12}}{\pi} \frac{\rho_{12}(s'_{12})G^0_L(s'_{12})}
 {s'_{12}- s_{12}}
( \langle A^0_{13}(s'_{13} \rangle_0 +
  \langle A^0_{23}(s'_{23} \rangle_0). \label{u11}
\ee
Here I exclude $\lam_S$ from the dispersion integral, but it is
also possible to include it: in both cases the unitarity
 is satisfied but with different behaviour of the
amplitude at the infinity, which cannot be defined by the
unitarity and analyticity only.  After the transition from one
rescattering to the full set of binary interactions in the final
state one has
 \ba A^0_{12}(s_{12})= \frac{\lam_S(s_{12})
 \;B_0(s_{12})}{1-B_0(s_{12})}\; + \nonumber \ea \ba
 \frac{1}{1-B_0(s_{12})}
  \int_{(m_1+m_2)^2}^{\infty}
 \frac{ds'_{12}}{\pi} \frac{\rho_{12}(s'_{12})G_0^L(s'_{12})}
 {s'_{12}- s_{12}}
( \langle A^0_{13}(s'_{13}) \rangle_0 + \langle
A^0_{23}(s'_{23}) \rangle_0) \label{u12}. \ea

Let us now check  that the extraction of the final state interaction
does not violate the unitarity condition (\ref{u5}). To calculate the
lhs of eq. (\ref{u5}),
the equation (\ref{u12}) is rewritten as follows:
\be
A^0_{12}(s_{12})=\frac{1-B_0^{*}(s_{12})}{\mid 1-B_0(s_{12}) \mid^{2}}
          \biggl( b^{\lam}(s_{12}) + J(s_{12}) \biggr),      \label{u13}
\ee
where
\be
b^{\lam}(s_{12}) = \lam_S(s_{12}) \; B_0(s_{12}),
                                     \label{u14}
\ee
\be
J(s_{12}) =
  \int_{(m_1+m_2)^2}^{\infty}
 \frac{ds'_{12}}{\pi} \frac{\rho_{12}(s'_{12})G_0^L(s'_{12})}
 {s'_{12}- s_{12}}
( \langle A^0_{13}(s'_{13}) \rangle_0
+ \langle A^0_{23}(s'_{23}) \rangle_0) \label{u15}.
\ee
 Thus,
 \ba
 disc_{12}\;A^0_{12}(s_{12})= \frac{1}{\mid 1-B_0(s_{12}) \mid^{2}} \biggl(
 Im\;B_0(s_{12})b^{\lam}(s_{12}) +
 \nonumber\\
 +(1-B_0(s_{12}))Im\;b^{\lam}(s_{12}) +
 Im\;B_0(s_{12}) J(s_{12}) +(1-B_0(s_{12}))disc\;J(s_{12})
 \biggr) \label{b16}.
  \ea
 Taking into account that
\ba
disc\;B_0(s_{12})= Im\;B_0(s_{12}) = \rho_{12}(s_{12})
G_0^L(s_{12}) , \nonumber \ea \be disc\;b^{\lam}(s_{12})=
\rho_{12}(s_{12})\; G_0^L(s_{12}) \lam_S, \ee \ba
disc\;J(s_{12})= \rho_{12}(s_{12}) G_0^L(s_{12}) \langle A^0_{13} + A^0_{23}
 \rangle_0,
\nonumber
\ea
I get the rhs of eq. (\ref{u5}), hence, the unitarity condition
is fulfilled.

\subsection{ P-wave interaction in the final state}

First, determine the structure of the amplitude,
where particles 1 and 2  interact in the P-wave and the
particle 3 is  spectator.
 The easiest way to do this is to transform the
decay amplitude  into  scattering amplitude. To perform this
transformation the antiparticle  3 with the
momentum $(-k_3)$ is considered instead of particle 3 itself.
The assumed form of the amplitude is
 $O_{1\mu} Q_{1\mu} A^1_{12}(s_{12})$.
Operator $Q_1$ describes the P-wave angular distribution of
particles 1 and 2 in the final state, it is defined as relative
momentum of these particles \be
Q_{1\mu}=k_{1\mu}-k_{2\mu}-\frac{m_1^2-m_2^2}{s_{12}}\;(k_1+k_2)_\mu.
\ee
Operator $O_1$ should be constructed as relative momentum of
initial state and antiparticle 3.  Taking into
account that $Q_{1\mu}\;(k_1+k_2)_\mu=0$, $O_1$ is defined as
 \be
O_{1\mu}= k_{3\mu}.
\ee
As is seen from eq. (\ref{u3}),
\be
F_1=O_{1\mu} Q_{1\mu},
\ee
and it is easy to find out that $F_1$ is proportional to $z_{13}$ in the
cms of particles 1 and 2.

 Hereafter  the procedure of the above section is used to
 write down the two-particle unitarity condition and the
integral equation for $A^1$.  The P-wave  two-particle
scattering amplitude of particles 1 and 2 in the N/D method can
be written as
 \ba A^1_{2\rightarrow 2}(s_{12})= Q_{1\mu}
 \;\frac{G_1^L(s_{12})}{1-B_1(s_{12})}\;Q_{1\mu}, \ea where the
B-function is equal to
 \be
  B_1(s_{ij})= \int_{(m_i+m_j)^2}^{\infty}
\frac{ds'}{\pi} \frac{G_1^L(s') \rho_{ij}(s') \langle Q_{1\mu}
Q_{1\mu} \rangle }{s'-s_{ij}},
 \ee
 and $\langle \dots \rangle$ means
the averaging over space angle.

As the first step, one should consider triangle diagram
with one P-wave rescattering of particles 1 and 2. The
discontinuity of this diagram is equal to:
  \ba
   disc_{12} A^1_{12}(s_{12})\;=\;
\frac{1}{2} \int d\Phi_{12}(k_1,k_2) a_{13}(s_{13},z_{13})
 Q_{1\mu} G_1^L(s_{12}) = \nonumber  \\ =\; G_1^L(s_{12})
\rho_{12}(s_{12}) \int \frac{d\Omega}{4\pi} Q_{1\mu} a_{13}(s_{13},z_{13}).
                                       \label{p2}
\ea
The integration over space angle is performed in the c.m. frame of particles
1 and 2. In this frame $Q_1$ turns  into $\vec k_{12}$. The
z-axis being directed along $\vec k_3$, the components of $\vec k_{12}$
are equal to
\ba
k_{12x}\;=\;k_{12}\;\sin\theta_{13}\;\cos\phi,\;\;\;
k_{12y}\;=\;k_{12}\;\sin\theta_{13}\;\sin\phi,\;\;\;
k_{12z}\;=\;k_{12}\;\cos\theta_{13}     \label{p2a}
\nonumber\\
k_{12}\;= \sqrt{
\;\frac{1}{s_{12}}\;[s_{12}-(m_1+m_2)^2][s_{12}-(m_1-m_2)^2]},
\ea
where $\phi$ is azimuthal angle. The integration over $\phi$
in eq. (\ref{p2}) keeps
the  components of  $Q_{1\mu}$ with $\mu=z$ only:
\ba
\int \frac{d\Omega}{4\pi} Q_{1x} \dots \;=\;
\int \frac{d\Omega}{4\pi} Q_{1y} \dots \;=\;0, \nonumber \\
\int \frac{d\Omega}{4\pi} Q_{1z} \dots \;=\;k_{12}
\int \frac{dz_{13}}{2}\; z_{13} \dots
\ea
Let us introduce the vector $k^\bot_3$ with only
 $z$ component unequal to zero in the cms:
\ba
k^{\bot}_{3\mu}=k_{3\mu}-p_\mu \frac{(p\;k_3)}{p^2}\,,
\ea
where $p=k_1+k_2$. Then it follows from eq. (\ref{p2})
\ba
disc\;=\; k^\bot_{3\mu}
\rho_{12}(s_{12}) G_1^L(s_{12}) \frac{k_{12}}{\sqrt{-(k^\bot_3)^2}}
\int_{C_3(s_{12})} \frac{dz_{13}}{2}\; z_{13}\;
a_{13}(s_{13},z_{13}), \label{p3} \ea where \ba
-(k^\bot_3)^2\;=\;\frac{[(M-m_3)^2-s_{12}][(M+m_3)^2-s_{12}]}{4s_{12}}.
\ea
The invariant part of the P-wave amplitude can be
written as a dispersion integral over the energy squared of particles
1 and 2 in the intermediate state, while $k^\bot_3$ in eq. (\ref{p3})
defines the operator structure of  the P-wave amplitude. Thus, the
P-wave triangle diagram is equal to
\ba
 k^\bot_{3\mu}  \int_{(m_1+m_2)^2}^{\infty}
 \frac{ds'_{12}}{\pi} \frac{\rho_{12}(s'_{12})G_1^L(s'_{12})}
 {s'_{12}- s_{12}}
 \langle a_{13} \rangle_1 ,   \label{p4}
\ea
where
\ba
 \langle a_{13} \rangle_1 \;=\; \frac{k_{12}}{\sqrt{-(k^\bot_3)^2}}\;
\int_{C_3(s_{12})} \frac{dz}{2} z\; a_{13}(s_{13},z).
\ea
Equation (\ref{p4}) should be multiplied by the operator $Q_1$ which describes
 angular distribution of
particles 1 and 2 in the final state. Using
$k^\bot_{3\mu}\;Q_{1\mu}= O_{1\mu}\;Q_{1\mu} $, one may
conclude that   the operator part of
$a_{12}$ is correctly reconstructed.
 To take into account binary rescattering  in
the final state it is necessary to multiply eq. (\ref{p4}) by
the factor $(1-B_1(s_{12}))^{-1}$.

The same steps should be done, if
$a_{13}$ is replaced by the direct
production term $\lam(s_{12},s_{13},s_{23})$.
Still, the energy dependence of $\lam$ is not usually known, so
use a simpler assumption is used.
 Let us introduce the constants
$\lam^P_{ij}$, which define the direct production amplitude of
particles $i$ and $j$ in the P-wave. Then  the following
integral equation for the amplitude $A_{12}^1$ can be written:
\ba A^1_{12}(s_{12})=
 \frac{\lam^P_{12} \;B_1(s_{12})}{1-B_1(s_{12})} +
\nonumber
\ea
\ba
 \frac{1}{1-B_1(s_{12})}
  \int_{(m_1+m_2)^2}^{\infty}
 \frac{ds'_{12}}{\pi} \frac{\rho_{12}(s'_{12})G_1^L(s'_{12})}
 {s'_{12}- s_{12}}
( \langle a_{13} \rangle_1 + \langle a_{23} \rangle_1 ).
                                                        \label{p5}
 \ea

\subsection{ D-wave interaction in the final state}

Likewise the case of D-wave interaction in the final state may be
investigated.  The amplitude, where particles 1 and 2 have the
last interaction in the D-wave, has a form $O_2 Q_2
A^2_{12}(s_{12})$. The operator $Q_2$ is a traceless tensor of
the range 2 constructed of relative momenta of particles 1 and
2:  \ba
Q_{2\mu\nu}\;=\;k_{12\mu}k_{12\nu}\;-\;\frac{1}{3}k_{12}^2 g^\bot_{\mu\nu},
\ea
where
\ba
k_{12\mu}\;=\;k_{1\mu}-k_{2\mu}-\frac{m_1^2-m_2^2}{s_{12}}\;(k_1+k_2)_\mu,
\ea                                                \nonumber
and
\ba
g^\bot_{\mu\nu}\;=\; g_{\mu\nu} - \frac{p_\mu p_\nu}{p^2}.
\ea                              \nonumber
Another operator, $O_2$, can be defined as
\ba
O_{2\mu\nu}\;=\;k_{3\mu}k_{3\nu}.
\ea
Below it will be proved that $O_2$ can be defined in this way.
Let us note here that the operator part of the D-wave decaying
amplitude is proportional to $3z_{13}^2-1$.

Let us define the D-wave scattering amplitude of particles 1 and 2 in
N/D representation as
\ba
A^2_{2\rightarrow 2}(s_{12})=
 Q_{2\mu\nu} \;\frac{G_2^L(s_{12})}{1-B_2(s_{12})}\;Q_{2\mu\nu}.
\ea
The discontinuity of the triangle diagram with the D-wave
rescattering of particles 1 and 2 is equal to
\ba
disc_{12}A^2_{12}(s_{12})\;=\;
 \frac{1}{2} \int d\Phi_{12}(k_1,k_2) a_{13}(s_{13},z_{13})
Q_{2\mu\nu} G_2^L(s_{12}) = \nonumber  \\
=\; G_2^L(s_{12})
\rho_{12}(s_{12}) \int \frac{d\Omega}{4\pi} Q_{2\mu\nu}
a_{13}(s_{13},z_{13}).
                                       \label{d1}
\ea
The integration over space angle is performed in the c.m. frame of
particles 1 and 2. Equation (\ref{p2a}) is used, and  only the
following components of tensor $k_{12\mu}k_{12\nu}$ are not equal to zero:
\ba
\int \frac{d\Omega}{4\pi}\;k_{12x}k_{12x}\dots\;=\;
\int \frac{d\Omega}{4\pi}\;k_{12y}k_{12y}\dots\;=\;
k^2_{12} \int \frac{dz_{13}}{4}\;(1-z_{13}^2)\dots  \nonumber \\
\int \frac{d\Omega}{4\pi}\;k_{12z}k_{12z}\dots\;=\;
k^2_{12} \int \frac{dz_{13}}{2}\;z_{13}^2\dots  \label{d2}
\ea
To perform the space integration in the invariant form,
let us define the tensor $g_{\mu\beta}^{\bot\bot}$ as
\be
g_{\mu\nu}^{\bot\bot}\;=\;g_{\mu\nu} - \frac{p_\mu p_\nu}{p^2}
-\frac{k^\bot_{\mu 3} k^\bot_{\nu 3}}{(k^\bot_3)^2}\,.
\ee
In the c.m. frame of particles 1 and 2 the tensor
$g_{\mu\nu}^{\bot\bot}$ has only two non-zero
diagonal elements:
\be
g_{\mu\beta}^{\bot\bot} \rar diag(0,-1,-1,0)\,,  \nonumber
\ee
Thus, eq. (\ref{d2}) can be rewritten as
\ba
\int \frac{d\Omega}{4\pi}\;k_{12\mu}k_{12\nu}\dots\;=\;
-g_{\mu\nu}^{\bot\bot}\;
k^2_{12} \int \frac{dz_{13}}{4}\;(1-z_{13}^2)\dots\;-  \nonumber \\
-\frac{k^\bot_{\mu 3} k^\bot_{\nu 3}}{(k^\bot_3)^2}\;
k^2_{12} \int \frac{dz_{13}}{2}\;z_{13}^2\dots         \label{d3}
\ea
Equation (\ref{d3}) should be placed into eq. (\ref{d1}) for the
discontinuity of triangle diagrams. Taking into account that
this expression must be multiplied by the external operator $Q_2$,
one can see that only the third term in eq. (\ref{d3})  contributes
into eq. (\ref{d1}). Moreover, one could replace $k^\bot_3$ by $k_3$.
Finally, as follows from eq. (\ref{d1}),
\ba
disc\;=\;    k_{3\mu}k_{3\nu}
\rho_{12}(s_{12})G_2^L(s_{12})
 \langle a_{13} \rangle_2 ,   \label{d4}
\ea
where
\ba
 \langle a_{13} \rangle_2 \;=\; \frac{k_{12}^2}{(k^\bot_3)^2}\;
\int_{C_3(s_{12})} \frac{dz}{4}(1-3z^2) \; a_{13}(s_{13},z).
\ea
The invariant part of the D-wave amplitude can be written as a dispersion
integral and a full set of binary rescatterings in the final
state are defined by the factor $(1-B_2(s_{12}))^{-1}$. As a
result, I come to the following integral  equation for the
amplitude $A_{12}^2$:
 \ba A^2_{12}(s_{12})= \frac{\lam^D_{12}
 \;B_2(s_{12})}{1-B_2(s_{12})} + \nonumber \ea \ba
 \frac{1}{1-B_2(s_{12})}
  \int_{(m_1+m_2)^2}^{\infty}
 \frac{ds'_{12}}{\pi} \frac{\rho_{12}(s'_{12})G_2^L(s'_{12})}
 {s'_{12}- s_{12}}
( \langle a_{13} \rangle_2 + \langle a_{23} \rangle_2 ),
                                                        \label{d5}
 \ea
where the constant $\lam^D_{ij}$ stands for the direct
production amplitude of particles $i$ and $j$ in the D-wave.

\section{Dispersion equations with resonance and non-resonance
production of particles}

In this Section  more realistic case is considered, when there
are resonance and non-resonance interactions between two
particles.  This situation happens, for example, in the $0^{++}$
wave of the pion-pion amplitude, where there is a non-resonance
background and a set of resonances at the energies above 1 GeV.

I start with the dispersion representation of the two-particle
amplitude for this particular case. As the simplest case, consider
the S-wave one-channel amplitude. The first resonance
term of the amplitude can be written as
\be
\sum_{\alpha} \frac{g^{(\alpha)2}(s)}{M^2_\alpha- s},
\ee
where $M_\alpha$ is a pure (non-physical) mass of resonance
$\alpha$, and the function $g^\alpha (s)$ describes its decay
into two particles. Generally, these functions can depend on
$s$, so the following form is suggested:
 \be g^\alpha (s) =
g^\alpha \phi(s).  \ee
 The second term of the amplitude with one
virtual loop is equal to
 \be \sum_{\alpha \alpha '}
\frac{g^{(\alpha)2}(s)}{M^2_\alpha- s} g^{(\alpha)} g^{(\alpha
       ')} b(s) \frac{g^{(\alpha ')2}(s)}{M^2_{\alpha '}- s},
\ee
where $b(s)$ is loop diagram with the cutoff function. The
behaviour of $\phi(s)$ at large $s$ (or at small distances $r$)
is not known, and the simplest way to avoid these uncertainties
is to assume that the contribution from $r < r_0$ is equal to
zero.  Then the cutoff function $\Lambda (s) $ is defined as
follows:
\be \Lambda
(s)= \int d^3r\;e^{i \vec k \vec r} \Theta(r-r_0) \int
         \frac{d^3k'}{(2\pi)^3}\;e^{-i \vec k' \vec r} \phi^2
         (s')    ,           \label{f0} \ee
          and $b(s)$ is equal to
 \be
  b(s)= i\rho(s) + P \int \frac{ds'}{\pi}\;
\frac{\rho(s')} {s'-s} \;\Lambda(s')
\label{f1_1}. \ee
 Summing up the terms with different
number of loops, one obtains
 the following expression for the amplitude
\be
A=\frac{\sum_{\alpha} \frac{g^{(\alpha)2}(s)}{M^2_\alpha- s}}
{1-b(s)\sum_{\alpha} \frac{g^{(\alpha)2}}{M^2_\alpha- s}}\; .
\label{f1}
\ee
 For non-resonance interaction of particles
 defined by vertex function $f(s)$, eq. (\ref{f1}) should be
rewritten as follows:
\be
A=\frac{\sum_{\alpha} \frac{g^{(\alpha)2}(s)}{M^2_\alpha- s}
+f(s)} {1- \biggl\{ b(s)\sum_{\alpha}
\frac{g^{(\alpha)2}}{M^2_\alpha- s} +b_f(s) \biggr\} }\;,
                                      \label{f2}
\ee
where
\be
b_f(s)= i\rho(s) + P \int \frac{ds'}{\pi}\; \frac{\rho(s')}
{s'-s} \;\Lambda_f(s') ,               \label{f1_2}
\ee
and $\Lambda_f(s)$ is given by eq. (\ref{f0}) with the replacement of
$\phi^2(s)$ by $f(s)$. It should be noted here that if $\phi$ and $f$
are  constant, then cutoff functions are equal to zero, and
dispersion representation coincides  with the K-matrix
 approach.

 Equation (\ref{f2}) can be easily generalized for the case of the
two-particle interaction with orbital momentum $L$:
\be
A_L=Q_{L,\mu}
\frac{\sum_{\alpha} \frac{g^{(\alpha)2}(s)}{M^2_\alpha- s}
+f(s)} {1- \biggl\{ b(s)\sum_{\alpha}
\frac{g^{(\alpha)2}}{M^2_\alpha- s} +b_f(s) \biggr\} } Q_{L,\mu},
\ee
where $b(s)$ and $b_f(s)$ are given by eqs.(\ref{f1_1}) and
(\ref{f1_2}), with the cutoff functions $\Lambda(s)$ and
$\Lambda_f(s)$, correspondingly, which are equal to
\be
\Lambda
(s)= \int d^3r\;e^{i \vec k \vec r} \Theta(r-r_0) \int
         \frac{d^3k'}{(2\pi)^3}\;e^{-i \vec k' \vec r} \phi^2
         (s') \langle Q_{L,\mu} Q_{L,\mu} \rangle ,
              \label{sss}
\ee
\be
\Lambda_f
(s)= \int d^3r\;e^{i \vec k \vec r} \Theta(r-r_0) \int
         \frac{d^3k'}{(2\pi)^3}\;e^{-i \vec k' \vec r} f
         (s') \langle Q_{L,\mu} Q_{L,\mu} \rangle
\ee

The same formalism can be applied to the case of multichannel
amplitude. The decay of resonance $\alpha$ into particles $m$
and $n$
is given by the function $g^{(\alpha)}_{mn}(s)=g^{(\alpha)}_{mn}
\phi_{mn}(s) $ and the
non-resonance transition from the channel with
particles $m$ and $n$ to the channel with particles $m'$ and $n'$ is
given by $f_{mn;m'n'}(s)$. Here the expression
for multichannel amplitude is not given in an
 explicit form, the recipe of
its construction can be found, for example, in \cite{Delta92}. I would like
only to note that the denominator in eq. (\ref{f2}), which
describes the rescattering of particles in multichannel case, has
 the matrix form $( \hat I-\hat B)^{-1}$, where $\hat I$ is a
 unit matrix and the B-matrix element is equal to \be
B_{mn;m'n'}= b_{mn;m'n'}(s) \sum_{\alpha}
\frac{g^{(\alpha)}_{mn}g^{(\alpha)}_{m'n'} }{M^2_\alpha- s} +
\beta_{mn;m'n'}(s)\;, \ee \be b_{mn;m'n'}(s)= i\rho_{mn}(s) + P
\int \frac{ds'}{\pi}\; \frac{\rho_{mn}(s')} {s'-s}
\;\Lambda_{mn;m'n'}(s'), \ee \be \beta_{mn;m'n'}(s)=
i\rho_{mn}(s) + P \int \frac{ds'}{\pi}\; \frac{\rho_{mn}(s')}
{s'-s} \;\Lambda_{f,mn;m'n'}(s').  \ee

Let us generalize the results of the previous sections and write
down the dispersion equations for the
three particle decay amplitude.
 The amplitude for the production of three
different particles  $k$, $m$, $n$, which is denoted as
$M_{kmn}$, is the following
 \be M_{kmn}\;=\; a_{km;n}(s_{12},z)+
a_{nm;k}(s_{13},z)+ a_{kn;m}(s_{23},z),
\ee
where the first term in the rhs gives the
amplitude with the last interaction between particles $k$ and $m$,
and so on. The interaction of two particles, for example, $n$ and
$m$, can happen in  different channels. Here I neglect the
isospin ( let $I=0$), so the two-particle momentum
 describes the type of interaction. Hence, the amplitude
 $a_{nm.k}$ can be written:
   \be a_{nm;k}= \sum_J F_J
 A^{0J}_{nm;k}\;, \ee
  where $A^{0J}_{nm;k}$ is the amplitude with
the last interaction of particles $n$ and $m$ with the momentum
$J$ and isospin $I=0$. The integral equation for $A^{0J}_{nm;k}$
can be written as a generalization of eqs.(\ref{u12}),
(\ref{p5}) and (\ref{d5}).  I start with the
term, where only particles $m$ and $n$ interact with each
other and the particle $k$ is spectator. The resonance and
non-resonance production of these particles from the initial
state is
 \be \lambda^{0J}_{mn;k}(s)= \sum_{\alpha}
\frac{C^{(\alpha)}_k g^{(\alpha)}_{mn}(s)} {M^2_\alpha -s} +
\Phi^{0J}_{mn;k}, \ee
 where $C^{(\alpha)}_k$ is the production
constant of resonance $\alpha$ and particle $k$,
$\Phi^{0J}_{mn;k}$ corresponds to the direct production of
particles $m$ and $n$ with relative momentum $J$. The
rescattering of particles $m$ and $n$ gives the following
amplitude:
\be
 \sum_{(n'm')} \lambda^{0J}_{n'm';k}
 \biggl\{ (\hat I - \hat B)^{-1} \biggr\}^{0J}_{n'm';nm},
 \ee
where the summation over intermediate states with the
production of $m'$ and $n'$ particles is performed.
The expression for triangle diagram can be written in the same
way. Finally, I get the following three-body dispersion
relation:
 \ba
 A^{0J}_{nm;k}\;=\;\sum_{(n'm')} \lambda^{0J}_{n'm';k}
 \biggl\{(\hat I - \hat B)^{-1} \biggr\}^{0J}_{n'm';nm}\;+
 \sum_{(n'm')} \sum_{(n"m")}
 \biggl\{ (\hat I - \hat B)^{-1} \biggr\}^{0J}_{n"m";nm}    \nonumber
 \ea
 \be
  \int
 \frac{ds'_{12}}{\pi}
 \frac{\rho_{m'n'}(s'_{12})
 N^{0J}_{m'n';m"n"}(s'_{12},s_{12})} {s'_{12}-
s_{12}} ( \langle a_{n'k;m'}(s'_{13},z) \rangle_J + \langle
a_{m'k;n'}(s'_{23},z) \rangle_J ),
\ee
where
\be
N^{0J}_{m'n';mn}(s'_{12},s_{12})= \Lambda_{f,m'n';mn} +
\sum_{\alpha} \frac{g^{(\alpha)}_{m'n'}
g^{(\alpha)}_{mn}(s)} {M^2_\alpha -s} \bar \Lambda(s')
\ee
and $\bar \Lambda(s')$ is the cutoff function obtained in eq. (\ref{sss})
with the help of
$\phi_{m'n'} (s)$.

\section{Dispersion equations for the coupled decay channels
$\bar p p$ ($0^{-+}$) $\rightarrow\;\pi^0 \pi^0 \pi^0,\; \eta
\pi^0 \pi^0,\; \eta \eta \pi^0, \; \bar K K \pi^0$}

The proton-antiproton annihilation in the state $J^{PC}=0^{-+}$
can originate from the two isospin states $I=0$ and $I=1$.
Due to isospin conservation in strong interactions there exist two
integral equations for the following decay channels:\\
$\bar p p$ ($IJ^{PC}=10^{-+}$) $\rightarrow\;\pi^0 \pi^0
\pi^0,\; \eta \eta \pi^0, \; \bar K K
\pi^0$\\
and\\
$\bar p p$ ($IJ^{PC}=00^{-+}$) $\rightarrow\;
 \eta \pi^0 \pi^0,\; \bar K K \pi^0$\\
 Let us write integral equations for the decay from
 $I=1$ state.

  \vspace{0.5cm}
  {\it a) Reaction $\bar p p$ ($10^{-+}$) $\rightarrow\;\pi^0
  \pi^0 \pi^0$}
  \vspace{0.5cm}

For this reaction the only difference  from the analysis
 given above is that one should take into account the isospin
 structure of the amplitude. The same analysis has been done in refs.
 \cite{Eta93}, \cite{Eta96}, where the unitarity condition in the
$\pi \pi$ channel was used to derive the amplitude for the $\eta
\rightarrow 3 \pi^0$ decay.  This decay goes with
violation of isospin
symmetry, so in both cases  initial states have the same
quantum numbers. Because of that, here the results of
refs. \cite{Eta93} and \cite{Eta96} are reproduced,
  and the contribution from the $\eta \eta$ and $\bar K K$ states
is also taken into account. The
annihilation amplitude into $3 \pi^0$ is
\be
M_{\pi^0 \pi^0
\pi^0}\;=\; a_{\pi^0 \pi^0; \pi^0}(s_{12},z) + a_{\pi^0 \pi^0;
\pi^0}(s_{13},z) + a_{\pi^0 \pi^0; \pi^0}(s_{23},z),
\ee
where
\be
a_{\pi^0 \pi^0; \pi^0}=a^0_{\pi \pi; \pi^0} +
\frac{4}{3} a^2_{\pi\pi; \pi^0},              \label{l41}
\ee
and $a^I_{\pi \pi; \pi^0}$ is the amplitude with the last
pions interacting in the isospin state $I$.
It should be noted that, due to the C-invariance,  there
is no term in eq. (\ref{l41}) with pion interactions in the state $I=1$ .
For the channel with isospin $I=0$  S- and D-wave interactions should be
taken into account; this allows  one to
 calculate properly the production of $f_0$
and $f_2$ resonances, so
 \be
 a^0_{\pi \pi; \pi^0}=A^{00}_{\pi \pi; \pi^0}
 + F_2 A^{02}_{\pi \pi; \pi^0}.
 \ee
 In the channel with isospin $I=2$
only S-wave interaction should be accounted for, therefore
 \be
 a^2_{\pi
 \pi; \pi^0}=A^{20}_{\pi \pi; \pi^0}.
 \ee
 Integral equations for
 $A^{0J}_{\pi \pi; \pi^0}$ have the following form:  \ba
 A^{0J}_{\pi \pi;
 \pi^0}\;= \; \lambda^{0J}_{\pi \pi; \pi^0} \biggl\{(\hat I - \hat B)^{-1}
 \biggr\}^{0J}_{\pi \pi; \pi \pi}\;+ \biggl\{ (\hat I - \hat B)^{-1}
 \biggr\}^{0J}_{\pi \pi; \pi \pi} \int \frac{ds'_{12}}{\pi} \frac{\rho_{\pi
 \pi}(s'_{12}) N^{0J}_{\pi \pi; \pi \pi}
 (s'_{12},s_{12})} {s'_{12}-s_{12}} \times
 \nonumber
 \ea
 \ba
  \times
 ( \langle \frac{2}{3} a^0_{\pi \pi; \pi^0}
 (s'_{13},z) \rangle_J +
 \langle \frac{20}{9} a^2_{\pi \pi; \pi^0}
 (s'_{23},z) \rangle_J +
 \langle \frac{4}{3} a^1_{\pi \pi; \pi^0}
 (s'_{23},z) \rangle_J )
  + \Delta_{\eta \eta} + \Delta_{\bar K K},   \label{l42}
  \ea
 where the contributions from the $\eta \eta$ and $\bar K K$
 intermediate states are:
 \ba
 \Delta_{\eta \eta} =
  \lambda^{0J}_{\eta \eta; \pi^0}
 \biggl\{(\hat I - \hat B)^{-1} \biggr\}^{0J}_{\eta \eta; \pi \pi}\;+
  \nonumber
  \ea
  \ba
 \biggl\{ (\hat I - \hat B)^{-1} \biggr\}^{0J}_{\eta \eta; \pi \pi}
    \int \frac{ds'_{12}}{\pi} \frac{\rho_{\eta
 \eta}(s'_{12}) N^{0J}_{\eta \eta; \pi \pi}
 (s'_{12},s_{12})} {s'_{12}-s_{12}}
  \langle 2\; a_{\pi^0 \eta; \eta}
 (s'_{13},z) \rangle_J \;,
 \ea
 \ba
 \Delta_{\bar K K} =
  \lambda^{0J}_{\bar K K; \pi^0}
 \biggl\{(\hat I - \hat B)^{-1} \biggr\}^{0J}_{\bar K K; \pi \pi}\;+
  \nonumber
  \ea
  \ba
 \biggl\{ (\hat I - \hat B)^{-1} \biggr\}^{0J}_{\bar K K; \pi \pi}
    \int \frac{ds'_{12}}{\pi} \frac{\rho_{\bar K K
 }(s'_{12}) N^{0J}_{\bar K K; \pi \pi}
 (s'_{12},s_{12})} {s'_{12}-s_{12}}
  \langle 2\; a_{\pi^0 K; \bar K}
 (s'_{13},z) \rangle_J \; .
 \ea
 The amplitude $a^1_{\pi \pi; \pi^0}$ in eq. ({\ref{l42})
 can be found only from the $\bar p p$ annihilation into charged
 pions, so in our approach it could be simply replaced by the
 direct production amplitude of $\rho^{+} \pi^{-}$.
 In the integral equation for  $A^{20}_{\pi \pi; \pi^0}$
  only $\pi \pi$ intermediate states are taken into account, so one has:
 \ba
 A^{20}_{\pi \pi; \pi^0}\;=
 \; \lambda^{20}_{\pi \pi; \pi}
 \biggl\{(\hat I - \hat B)^{-1} \biggr\}^{20}_{\pi \pi;
 \pi \pi}\;+ \biggl\{ (\hat I - \hat B)^{-1}
 \biggr\}^{20}_{\pi \pi; \pi \pi}   \times
  \nonumber \ea \ba
  \times
    \int \frac{ds'_{12}}{\pi} \frac{\rho_{\pi \pi}(s'_{12})
 N^{20}_{\pi \pi; \pi \pi} (s'_{12},s_{12})}
 {s'_{12}-s_{12}} ( \langle a^0_{\pi \pi; \pi^0} (s'_{13},z)
 \rangle_0 + \langle \frac{1}{3} a^2_{\pi \pi; \pi^0}
 (s'_{23},z) \rangle_0 +
 \langle a^1_{\pi \pi; \pi^0}
 (s'_{23},z) \rangle_0 )
  \ea
  \vspace{0.5cm}
  {\it b) Reactions $\bar p p$ ($0^{-+}$) $\rightarrow\;\eta
  \eta \pi^0$}
  \vspace{0.5cm}

  In the $\eta \pi^0$ channel the  S- and D-wave
  interaction with the production of $a_0$ and $a_2$ resonances
 are taken into consideration.
  The annihilation amplitude is the following:
  \be
M_{\eta \eta \pi^0}\;=\; a_{\eta \eta; \pi^0}(s_{12},z) +
a_{\eta \pi^0; \eta}(s_{13},z) + a_{\eta \pi^0;
\eta}(s_{23},z),\;
\ee
where
\be
a_{\eta \eta; \pi^0}=A^{00}_{\eta \eta; \pi^0}\;+
                         F_2 A^{02}_{\eta \eta; \pi^0},
\ee
\be
   a_{\eta \pi^0; \eta}= A^{10}_{\eta \pi^0; \eta} +
                         F_2 A^{12}_{\eta \pi^0; \eta}.
   \ee
   The integral equation for the $A^{0J}_{\eta \eta; \pi^0}$
   amplitude has the following form:
 \ba
 A^{0J}_{\eta \eta; \pi^0}\;=
 \; \lambda^{0J}_{\pi \pi; \pi^0}
 \biggl\{(\hat I - \hat B)^{-1} \biggr\}^{0J}_{\pi \pi; \eta
 \eta}\;+ \biggl\{ (\hat I  -\hat B)^{-1} \biggr\}^{0J}_{\pi
 \pi; \eta \eta}  \times
  \nonumber \ea
  \ba \times
  \int \frac{ds'_{12}}{\pi}
    \frac{\rho_{\pi \pi}(s'_{12}) N^{0J}_{\pi \pi; \eta \eta}
 (s'_{12},s_{12})} {s'_{12}-s_{12}}
 ( \langle \frac{2}{3} a^0_{\pi \pi; \pi^0}
 (s'_{13},z) \rangle_J +
 \langle \frac{20}{9} a^2_{\pi \pi; \pi^0}
 (s'_{23},z) \rangle_J +
 \ea
 \ba
 \langle \frac{4}{3} a^1_{\pi \pi; \pi^0}
 (s'_{23},z) \rangle_J )
  + \Delta_{\eta \eta} + \Delta_{\bar K
 K},
 \nonumber
  \ea
 where  contributions from the $\eta \eta$ and $\bar K K$
 intermediate states are:
 \ba
 \Delta_{\eta \eta} =
  \lambda^{0J}_{\eta \eta; \pi^0}
 \biggl\{(\hat I - \hat B)^{-1} \biggr\}^{0J}_{\eta \eta; \eta
 \eta}\;+  \nonumber \ea \ba
  \biggl\{ (\hat I - \hat B)^{-1} \biggr\}^{0J}_{\eta
 \eta; \eta \eta}  \int \frac{ds'_{12}}{\pi}
    \frac{\rho_{\eta \eta}(s'_{12}) N^{0J}_{\eta \eta; \eta
 \eta} (s'_{12},s_{12})} {s'_{12}-s_{12}} \langle 2\; a_{\pi^0
  \eta; \eta} (s'_{13},z) \rangle_J \;, \ea
   \ba
    \Delta_{\bar K
 K} = \lambda^{0J}_{\bar K K; \pi^0} \biggl\{(\hat - I \hat
 B)^{-1} \biggr\}^{0J}_{\bar K K; \eta \eta}\;+
  \nonumber   \ea \ba
   \biggl\{ (\hat I
 - \hat B)^{-1} \biggr\}^{0J}_{\bar K K; \eta \eta}
   \int \frac{ds'_{12}}{\pi} \frac{\rho_{\bar K K
 }(s'_{12}) N^{0J}_{\bar K K; \eta \eta}
 (s'_{12},s_{12})} {s'_{12}-s_{12}}
  \langle 2\; a_{\pi^0 K; K}
 (s'_{13},z) \rangle_J \;.
 \ea
   The integral equation for the $A^{1J}_{\eta \pi^0; \eta}$
   amplitude is:
  \ba
  A^{1J}_{\eta \pi^0; \eta} =
  \lambda^{1J}_{\eta \pi^0; \eta}
 \biggl\{(\hat - I \hat B)^{-1} \biggr\}^{1J}_{\eta \pi^0; \eta
 \pi^0}\;+ \nonumber \ea \ba \biggl\{ (\hat I - \hat B)^{-1}
 \biggr\}^{1J}_{\eta \pi^0; \eta \pi^0} \int
    \frac{ds'_{12}}{\pi} \frac{\rho_{\eta \pi^0}(s'_{12})
 N^{1J}_{\eta \pi^0; \eta \pi^0} (s'_{12},s_{12})}
 {s'_{12}-s_{12}} (\langle  a_{\eta \eta; \pi^0} (s'_{13},z)
 \rangle_J + \langle  a_{\eta \pi^0; \eta} (s'_{23},z)
 \rangle_J) \;.  \ea

  \vspace{0.5cm}
  {\it b) Reactions $\bar p p$ ($0^{-+}$) $\rightarrow\;\bar K
  K \pi^0$}
  \vspace{0.5cm}

These annihilation amplitudes are:
  \be
M_{\pi^0 \bar K K}\;=\; a_{\pi^0 K; \bar K}(s_{12},z) +
a_{\pi^0 \bar K; K}(s_{13},z) + a_{\bar K K;
\pi^0}(s_{23},z)\;,
\ee

As before, in the $\bar K K$ channel
the S- and D-wave interactions and
 $K^*$ resonances in the $K \pi$ channel are accounted for. So, one has:
\be
a_{\bar K K; \pi^0}=A^{00}_{\bar K K; \pi^0}\;+
                         F_2 A^{02}_{\bar K K; \eta},
\ee
\be
a_{\bar K \pi^0; K}=A^{1/21}_{\bar K \pi^0;K}\;
\ee
 For these amplitudes one can get the following integral
 equations:
  \ba
  A^{1/21}_{\bar K \pi^0; K} =
  \lambda^{1/21}_{\bar K \pi^0; K}
 \biggl\{(\hat I - \hat B)^{-1} \biggr\}^{1/21}_{\pi^0 K; \pi^0
 K}\;+ \biggl\{ (\hat I - \hat B)^{-1} \biggr\}^{1/21}_{\pi^0 K;
 \pi^0 K} \times \nonumber \ea \ba \times \int
    \frac{ds'_{12}}{\pi} \frac{\rho_{ \pi^0 K}(s'_{12}) N^{1/2
 1}_{\pi^0 K; \pi^0 K} (s'_{12},s_{12})} {s'_{12}-s_{12}}
  (\langle  a_{\bar K \pi^0; K}
 (s'_{13},z) \rangle_J +
  \langle  a_{\bar K K; \pi^0}
 (s'_{23},z) \rangle_J) \;.
 \ea
   The integral equation for the $A^{0J}_{\bar K K; \pi^0}$
   amplitude has the following form:
 \ba
 A^{0J}_{\bar K K; \pi^0}\;=
 \; \lambda^{0J}_{\pi \pi; \pi^0}
 \biggl\{(\hat I - \hat B)^{-1} \biggr\}^{0J}_{\pi \pi; \bar K
 K}\;+ \biggl\{ (\hat I - \hat B)^{-1} \biggr\}^{0J}_{\pi \pi;
 \bar K K} \times \nonumber \ea \ba \times \int
    \frac{ds'_{12}}{\pi} \frac{\rho_{\pi \pi}(s'_{12})
 N^{0J}_{\pi \pi; \bar K K} (s'_{12},s_{12})} {s'_{12}-s_{12}} (
 \langle \frac{2}{3} a^0_{\pi \pi; \pi^0} (s'_{13},z) \rangle_J
 + \langle \frac{20}{9} a^2_{\pi \pi; \pi^0} (s'_{23},z)
 \rangle_J + \ea \ba \langle \frac{4}{3} a^1_{\pi \pi; \pi^0}
 (s'_{23},z) \rangle_J )
  + \Delta_{\eta \eta} + \Delta_{\bar K
 K},
 \nonumber
  \ea
 where  contributions from the $\eta \eta$ and $\bar K K$
 intermediate states are as follows:
 \ba
 \Delta_{\eta \eta} =
  \lambda^{0J}_{\eta \eta; \pi^0}
 \biggl\{(\hat I - \hat B)^{-1} \biggr\}^{0J}_{\eta \eta; \bar K
 K}\;+  \nonumber \ea \ba
  \biggl\{ (\hat I - \hat B)^{-1} \biggr\}^{0J}_{\eta \eta;
 \bar K K}  \int \frac{ds'_{12}}{\pi}
    \frac{\rho_{\eta \eta}(s'_{12}) N^{0J}_{\eta \eta; \eta
 \eta} (s'_{12},s_{12})} {s'_{12}-s_{12}} \langle 2\; a_{\pi^0
  \eta; \eta} (s'_{13},z) \rangle_J \;, \ea
   \ba
    \Delta_{\bar K
 K} = \lambda^{0J}_{\bar K K; \pi^0} \biggl\{(\hat I  - \hat
 B)^{-1} \biggr\}^{0J}_{\bar K K; \bar K K}\;+
  \nonumber \ea
  \ba
   \biggl\{ (\hat I
 - \hat B)^{-1} \biggr\}^{0J}_{\bar K K; \bar K K}
    \int \frac{ds'_{12}}{\pi} \frac{\rho_{\bar K K
 }(s'_{12}) N^{0J}_{\bar K K; \bar K K}
 (s'_{12},s_{12})} {s'_{12}-s_{12}}
  \langle 2\; a_{\pi^0 K; K}
 (s'_{13},z) \rangle_J \;,
 \ea
A set of integral equations for the reactions
$\bar p p$ ($IJ^{PC}=00^{-+}$) $\rightarrow\;
 \eta \pi^0 \pi^0,\; \bar K K \pi^0$\\ may be written in the
 same way.

\section{Conclusion}
To summarize,  the two-particle discontinuity of the decay amplitude
is written, and
the integral equation is obtained which takes into
 account not only the S-wave binary interactions but also
 interactions with higher angular momenta (P- and D-wave).
 These equations are generalized for the resonance and
 non-resonance type of two-particle interactions. A complete
 three-body dispersion equation is derived, which may be used
 for the analysis of various three-particle reactions.  I
 demonstrate how the coupled three-body equations can be
written for the $\pi^0 \pi^0 \pi^0,\;\; \eta \pi^0 \pi^0,\;\;
\eta \eta \pi^0, \;\; \bar K K \pi^0$ channels in the $\bar p p$
annihilation at rest.

 This work is supported by the grants INTAS-93-0283 and RFFI
 96-02-17934.

\end{document}